\documentclass[11pt]{article}
\usepackage{graphicx}
\usepackage[margin=3cm]{geometry}
\usepackage{amsmath}
\usepackage{amssymb}
\usepackage{biblatex}
\usepackage{hyperref}
\usepackage{subcaption}
\usepackage{bbm}
\usepackage{placeins}

\title{Concentration of Stationary Measures onto Large Scales in the Fast-Advection Limit of the Stochastically Forced Two-Dimensional Navier--Stokes Equations}

\author{Tobias Rohner\\
\small Seminar for Applied Mathematics, ETH Zurich, CH-8092 Zurich, Switzerland\\
\texttt{tobias.rohner@sam.math.ethz.ch}}

\date{\today}

\begin{document}

\maketitle

\begin{abstract}
We investigate the fast-advection limit ($\varepsilon \to 0$) of the stochastically forced two-dimensional incompressible Navier--Stokes equations on the torus. Numerical simulations are performed at resolution $N = 128$ for four different forcing shells on both the square torus and a thin torus, with the advection parameter $\varepsilon$ varied over approximately one decade. We find that the stationary distributions become increasingly concentrated on the lowest Fourier modes as $\varepsilon \to 0$, with the energy-to-enstrophy ratio $E/\Omega$ and the fraction $R$ of enstrophy in the lowest modes both increasing monotonically. At finite $\varepsilon$ the degree of concentration depends on the forcing shell through the effective spectral value $B_1/B_0$, in agreement with the condensation bound of Sznitman and Widmayer. Linear extrapolation to $\varepsilon = 0$ suggests that the limiting values may be forcing-independent, with $B_1/B_0$ governing only the rate of convergence. The same qualitative behaviour persists on the thin torus.
\end{abstract}

\section{Introduction}

Two-dimensional turbulence differs fundamentally from its three-dimensional counterpart through the simultaneous conservation of kinetic energy and enstrophy in the inviscid limit. As first described by Kraichnan \cite{Kraichnan1967}, these conservation laws give rise to the dual-cascade scenario in which enstrophy is transferred towards small scales while energy is transported upscale through an inverse cascade. On finite domains the inverse cascade is arrested at the largest available scales, leading to an accumulation of energy in the lowest Fourier modes and the formation of large-scale coherent structures, commonly referred to as condensates \cite{Smith_Yakhot_1993, Boffetta_Ecke_2012}.

The onset of condensation has been studied extensively over the past decades. In particular, Linkmann, Hohmann and Eckhardt \cite{Linkmann_Hohmann_Eckhardt_2020} investigated the transition to two-dimensional turbulence under different forcing mechanisms and demonstrated that this transition is non-universal. They showed that stochastic forcing with prescribed energy injection exhibits a continuous transition to the condensate state, whereas forcing through a linear instability leads to a discontinuous transition accompanied by hysteresis. Their work established that the nature of condensate formation depends strongly on the forcing mechanism.

From the theoretical side, Sznitman and Widmayer \cite{SznitmanWidmayer2026a, SznitmanWidmayer2026b} recently established a rigorous framework for the inviscid limit of a Galerkin approximation to the stochastically forced Navier--Stokes equations with a small random stirring, playing the role of a regularization term. In \cite{SznitmanWidmayer2026a} they construct an effective energy--enstrophy diffusion process in a two-dimensional cone and prove a condensation bound: the ratio of expected energy to expected enstrophy is controlled by the effective spectral value $B_1/B_0$ of the forcing, so that forcings with small $B_1/B_0$ relative to the highest resolved wavenumber yield strong concentration on the lowest modes. In the companion article \cite{SznitmanWidmayer2026b} this effective diffusion is shown to arise as the inviscid limit of the stationary Galerkin process. These results provide quantitative predictions for the rate of condensation, while leaving open the questions of a possible limiting motion and what the role of the small stirring is.

The present work addresses a complementary problem. Rather than investigating how condensation emerges as a control parameter is varied, we study the asymptotic structure of statistically stationary condensates after they have formed. Specifically, we consider the fast-advection limit of the stochastically forced two-dimensional Navier--Stokes equations and ask whether the corresponding stationary distributions become increasingly concentrated on the lowest Fourier modes as the nonlinear dynamics becomes dominant. All simulations are evolved until approximate statistical stationarity, allowing us to investigate the limiting invariant measure rather than transient dynamics or the onset of a phase transition. The numerical findings are compared with the predictions of the Sznitman--Widmayer condensation bound.

We consider statistically stationary solutions of the Navier--Stokes equations on the two-dimensional square torus
\begin{equation*}
    \mathbb{T}=(\mathbb{R}/2\pi\mathbb{Z})\times(\mathbb{R}/2\pi\mathbb{Z})
\end{equation*}
with smooth additive stochastic forcing in the framework of Kuksin--Shirikyan \cite{Kuksin2012},
\begin{equation} \label{eq:snse}
    \mathrm{d}\omega + \frac1\varepsilon \mathbf u\cdot\nabla\omega\,\mathrm dt = \Delta\omega\,\mathrm dt + \sum_{\mathbf k} b_{\mathbf k}\varphi_{\mathbf k}\,\mathrm d\beta_{\mathbf k},
\end{equation}
subject to the initial condition
\begin{equation*}
    \omega|_{t=0}=0.
\end{equation*}

Here, $\omega$ is the voritcity, $\mathbf u = \nabla^\bot(-\Delta)^{-1}\omega$ is the velocity, and $\varphi_k$ are pairwise orthogonal eigenfunctions of the Laplacian satisfying
\begin{equation*}
    \|\varphi_k\|_{-1}^2 = 1, \qquad \int_{\mathbb T}\varphi_k\,\mathrm{d}z = 0.
\end{equation*}
the processes $\beta_k$ are independent Brownian motions, and only finitely many forcing coefficients $b_k$ are non-zero.

Following Kuksin and Shirikyan \cite{Kuksin2012} (See also \cite{HairerMattingly2006}), we choose the forcing according to
\begin{equation}
    \begin{aligned}
        \mathbb{Z}_+^2 &= \left\{ (k_1,k_2) \in \mathbb{Z}^2 \middle| k_1 > 0 \right\} \cup \left\{ (0,k_2) \in \mathbb{Z}^2 \middle| k_2 > 0 \right\}, \\
        \mathbb{Z}_-^2 &= \left\{ (k_1,k_2) \in \mathbb{Z}^2 \middle| -\mathbf{k} \in \mathbb{Z}_+^2 \right\}.
    \end{aligned}
\end{equation}
We set for $k \in \mathbb{Z}^2\setminus\left\{(0,0)\right\}$
\begin{equation}
    \varphi_k(z) = \begin{cases}
        \frac{|k|}{\sqrt{2}\pi}\cos(k\cdot z) &\mbox{ if } k \in \mathbb{Z}_+^2 \\
        -\frac{|k|}{\sqrt{2}\pi}\sin(k\cdot z) &\mbox{ if } k \in \mathbb{Z}_-^2
    \end{cases}.
\end{equation}
The coefficients $b_k$ are chosen in compliance with
\begin{equation}
    b_{\mathbf k} = 0 \mbox{ where } |\mathbf{k}|_{\infty} < k_{\min} \mbox{ or } k_{\max} \leq |\mathbf{k}|_{\infty},
\end{equation}
where $|\mathbf{k}|_{\infty} = \max(|k_1|,|k_2|)$ denotes the maximum norm.
Then, we have that \cite{Kuksin2012}
\begin{equation}
    \begin{aligned}
        \mathbb{E}\left[\Omega\right] &= \frac{B_0}{2}, \;\mbox{ where }\; B_0 = \sum_kb_k^2\left\|\varphi_k\right\|_{-1}^2 = \sum_kb_k^2, \\
        \mathbb{E}\left[\left\|\nabla\omega\right\|^2\right] &= \frac{B_1}{2}, \;\mbox{ where }\; B_1 = \sum_kb_k^2\left\|\varphi_k\right\|^2 = \sum_k|\mathbf{k}|^2b_k^2.
    \end{aligned}
\end{equation}

Writing the vorticity field in Fourier series as
$\omega = \sum_{\mathbf{k}} \omega_{\mathbf{k}}\, e^{i\mathbf{k}\cdot z}$,
we define the kinetic energy
\begin{equation*}
    E = \int_{\mathbb T} |\mathbf u|^2\,\mathrm{d}z = (2\pi)^2 \sum_{\mathbf{k}} |\mathbf{k}|^{-2}\,|\omega_{\mathbf{k}}|^2,
\end{equation*}
the enstrophy
\begin{equation*}
    \Omega = \int_{\mathbb T} |\omega|^2\,\mathrm{d}z = (2\pi)^2 \sum_{\mathbf{k}} |\omega_{\mathbf{k}}|^2,
\end{equation*}
and the energy-to-enstrophy ratio $E/\Omega$. Note that the factor $(2\pi)^2$ cancels in the ratio, so
$E/\Omega = \sum_{\mathbf{k}}|\mathbf{k}|^{-2}|\omega_{\mathbf{k}}|^2 / \sum_{\mathbf{k}}|\omega_{\mathbf{k}}|^2$.
To quantify the concentration of the stationary distribution on the lowest Fourier modes we further introduce the fraction
\begin{equation*}
    R = \frac{\sum_{|\mathbf{k}|=1}|\omega_{\mathbf{k}}|^2}{\sum_{\mathbf{k}}|\omega_{\mathbf{k}}|^2}.
\end{equation*}
of enstrophy contained in the modes with $|\mathbf{k}|=1$, where $|\mathbf{k}| = (k_1^2+k_2^2)^{1/2}$ denotes the Euclidean norm.

The principal questions addressed in this paper are:

\begin{enumerate}
    \item Do the quantities $\frac{E}{\Omega}$ and $R$ converge to positive limits as $\varepsilon \to 0$?
    \item Does the stationary distribution concentrate onto the lowest Fourier modes as $\varepsilon\to0$?
    \item Is the limiting behaviour independent of the forcing shell?
    \item How do the variances of global quantities and individual low Fourier modes behave in the fast-advection limit?
    \item Does the corresponding behaviour persist on thin tori?
\end{enumerate}

\section{Numerical Method}

All simulations are performed using the high-performance fluid solver \emph{azeban} \cite{rohner2024}, a GPU-accelerated pseudospectral code for incompressible flow on periodic domains, and the data is publicly available at \cite{rohner_2026_22895665}. The solver works with the velocity field in Fourier space, where incompressibility is enforced by the Leray projection $\mathbbm{1} - \mathbf{k}\mathbf{k}^\top/|\mathbf{k}|^2$ applied mode-by-mode, eliminating the need to solve a pressure equation. The nonlinear advection term is computed with the standard $2/3$-dealiasing rule, and fast Fourier transforms are performed using cuFFT on the GPU. We refer to \cite{rohner2024} for a detailed description of the solver architecture, parallelization strategy, and performance benchmarks.

\subsection{Stochastic forcing}

For the present work, \emph{azeban} was extended to support additive stochastic forcing. Although the equation~\eqref{eq:snse} is written in vorticity form, the solver operates on the velocity variables. The vorticity forcing $\sum_{\mathbf{k}} b_{\mathbf{k}}\varphi_{\mathbf{k}}\,\mathrm{d}\beta_{\mathbf{k}}$ is therefore translated into an equivalent velocity forcing by applying the Biot--Savart operator $\nabla^{\perp}(-\Delta)^{-1}$, which in Fourier space amounts to multiplying by $i\,\mathbf{k}^{\perp}/|\mathbf{k}|^2$ with $\mathbf{k}^{\perp} = (k_2, -k_1)$. Since the forcing eigenfunctions $\varphi_{\mathbf{k}}$ are normalized so that $\|\varphi_{\mathbf{k}}\|_{-1}^2=1$ and thus carry a factor of $|\mathbf{k}|$, the combined effect of the Biot--Savart operator and the eigenfunction normalization is a multiplication by $\mathbf{k}^{\perp}/|\mathbf{k}|$ in the implementation.

The forcing coefficients are chosen as
\begin{equation} \label{eq:forcing}
    b_{\mathbf{k}} = \begin{cases}
        C, & \mathbf{k}\in\mathbb{Z}_+^2 \text{ and } k_{\min}\leq |\mathbf{k}|_\infty < k_{\max},\\ -C, & \mathbf{k}\in\mathbb{Z}_-^2 \text{ and } k_{\min}\leq |\mathbf{k}|_\infty < k_{\max},\\
        0, & \text{otherwise},
    \end{cases}
\end{equation}
where
\begin{equation*}
    C = \sqrt{\frac{B_0}{(2k_{\max}-1)^2-(2k_{\min}-1)^2}}.
\end{equation*}
This normalization ensures that the prescribed energy injection rate $B_0$ remains constant when varying the forcing shell. At each timestep, independent standard normal random variables $\eta_{\mathbf{k}}, \nu_{\mathbf{k}} \sim \mathcal{N}(0,1)$ are generated for each forced mode, and the stochastic velocity increment in Fourier space takes the form
\begin{equation}
    \Delta\hat{\mathbf{f}}_{\mathbf{k}} = \frac{\sqrt{\varepsilon\,\Delta t}}{2\pi}\,\frac{C}{|\mathbf{k}|}\,
    \begin{bmatrix}
        k_2 \\
        -k_1
    \end{bmatrix}\,\bigl(\eta_{\mathbf{k}} + S_{\mathbf{k}}\, i\,\nu_{\mathbf{k}}\bigr),
\end{equation}
where $S_{\mathbf{k}} = +1$ for $\mathbf{k}\in\mathbb{Z}_+^2$ and $S_{\mathbf{k}} = -1$ for $\mathbf{k}\in\mathbb{Z}_-^2$. The factor $\sqrt{\varepsilon\,\Delta t}$ arises from the time rescaling $\tau = t/\varepsilon$ implicit in equation~\eqref{eq:snse}: the Brownian increments scale as $\Delta\beta \sim \sqrt{\varepsilon\,\Delta t}$ in the rescaled time variable. The same parameter $\varepsilon$ thus controls the advection strength ($1/\varepsilon$), the viscosity ($\varepsilon_N$ below), and the forcing amplitude ($\sqrt{\varepsilon}$).

\subsection{Viscosity}

The dissipation is implemented through the spectral viscosity framework of Tadmor \cite{Tadmor1989}. The viscosity operator acts in Fourier space as
\begin{equation*}
    -\varepsilon_N\,|\mathbf{k}|^{2s}\,\hat Q_k\,\hat{\mathbf{u}}_{\mathbf{k}},
\end{equation*}
where $s \geq 1$ is the hyperviscosity exponent, $\hat Q_k \in [0,1]$ is a Fourier-space damping profile, and $\varepsilon_N$ is a resolution-dependent viscosity coefficient. In this work we use $s = 1$ (standard Laplacian viscosity) with a step profile $\hat Q_k = \mathbf{1}_{\{|\mathbf{k}| > m_N\}}$, where $m_N = m_0\,N^{1/2}$ with $m_0 = 10^{-8}$. Since $m_N$ is negligible compared to all nonzero wavenumbers, $\hat Q_k = 1$ for all $\mathbf{k} \neq 0$, so the spectral viscosity reduces to standard Laplacian dissipation acting on all nonzero modes. The effective viscosity is $\varepsilon_N |\mathbf{k}|^2$ with
\begin{equation*}
    \varepsilon_N = \frac{\varepsilon}{N^{2s-1}} = \frac{\varepsilon}{N}.
\end{equation*}
The $1/N$ scaling is consistent with the convergence theory for spectral viscosity methods \cite{Tadmor1989, Lanthaler2020} and ensures that the dissipation vanishes as $N\to\infty$.

\subsection{Timestepping}

Because the stochastic forcing is additive, each Fourier mode evolves according to an SDE with diffusion coefficient independent of the state variable. Consequently, the Euler--Maruyama discretization is equivalent to the Milstein scheme, yielding strong order one convergence \cite{Kloeden2013}. The velocity equation is integrated by a single Euler--Maruyama step per timestep:
\begin{equation*}
    \hat{\mathbf{u}}^{n+1} = \hat{\mathbf{u}}^n + \Delta t\,\mathrm{d}\hat{\mathbf{u}}/\mathrm{d}t\big|_{t = t_n},
\end{equation*}
where the right-hand side is evaluated at the current state. The stochastic increment is generated independently at every timestep and added directly in Fourier space. The resulting fully discrete update for each Fourier mode is given by
\begin{equation}
    \hat{\mathbf u}^{n+1}_{\mathbf k} = \hat{\mathbf u}^{n}_{\mathbf k} - \left(\mathbbm 1 - \frac{\mathbf k\mathbf k^\top}{|\mathbf k|^2}\right)\left(i\mathbf k^\top\hat{\mathbf B}_{\mathbf k}^n \Delta t + \Delta\hat{\mathbf f}_{\mathbf k}\right) - \varepsilon_N |\mathbf k|^2 \hat Q_k \, \hat{\mathbf u}_{\mathbf k}^n \Delta t.
\end{equation}
Here $\hat{\mathbf{B}}_{\mathbf{k}}^n$ denotes the Fourier transform of the dealiased advection tensor $\mathbf{u}\otimes\mathbf{u}$, $\Delta\hat{\mathbf{f}}_{\mathbf{k}}$ is the stochastic forcing increment, and the zero mode $\mathbf{k} = 0$ is held fixed at zero.

The timestep is constrained by the advective and viscous CFL conditions,
\begin{equation*}
    \Delta t \leq C\,\min\!\left(\frac{1}{N\,|\mathbf{u}|_\infty},\; \frac{2}{\varepsilon_N\,(\pi N)^2}\right),
\end{equation*}
with a safety factor $C = 0.001$, and $\varepsilon_N = \varepsilon/N$ as above.

\subsection{Thin torus}

Simulations on the thin torus $\mathbb{T}_\delta = (\mathbb{R}/2\pi\delta^{-1}\mathbb{Z})\times(\mathbb{R}/2\pi\mathbb{Z})$ with $\delta \in \mathbb N$ are implemented by restricting the stochastic forcing to the sublattice of Fourier modes with $k_1$ divisible by $\delta$. Since the initial condition is zero, the nonlinearity preserves this sublattice, and the viscosity is diagonal in Fourier space, the solution remains confined to the sublattice for all time. Modes with $k_1 = \delta\,n$ correspond to functions with period $2\pi/\delta$ in~$x$, which is exactly the thin torus geometry. The per-mode forcing amplitude is rescaled by the ratio of the total number of shell modes to the number of sublattice modes, so as to compensate for the reduced number of active degrees of freedom on the sublattice. In this work we use $\delta = 2$, corresponding to $\mathbb{T} = (\mathbb{R}/\pi\mathbb{Z})\times(\mathbb{R}/2\pi\mathbb{Z})$.

\section{Results}

\subsection{Square Torus}

We simulate the stochastically forced Navier--Stokes equations for parameter combinations $(k_{\min}, k_{\max}) \in \{(10,11),\,(10,16),\,(20,21),\,(20,26)\}$ at a resolution of $N = 128$ on the standard two-dimensional torus $\mathbb{T} = (\mathbb{R}/2\pi\mathbb{Z})\times(\mathbb{R}/2\pi\mathbb{Z})$. The forcing strength is set to $B_0 = 8000$ and $\varepsilon$ is varied over the values $\varepsilon_i \in \bigl\{\frac{0.05}{N}\cdot 2^{-i/2}\bigr\}_{i=1}^{8}$, spanning roughly one decade from $\varepsilon_1 \approx 2.8\times10^{-4}$ to $\varepsilon_8 \approx 2.4\times10^{-5}$. The four forcing shells are chosen so that the effective spectral value $B_1/B_0$ of the forcing ranges from approximately $134$ to $687$ (see Table~\ref{tab:extrapolation}), allowing us to probe how the rate and degree of condensation depend on the forcing location. Snapshots of the energy spectrum, enstrophy spectrum, and low-resolution flow fields are stored up to time $T = 4\pi^2\cdot 3'000$ at $30'000$ equispaced timepoints. Unless otherwise stated, stationary averages are computed over the last $1/16$-th of the simulation time, by which point the global quantities have reached a statistically-stationary state.

\subsubsection{Concentration onto large scales}

\begin{figure}[ht!]
    \centering
    \includegraphics[width=\textwidth]{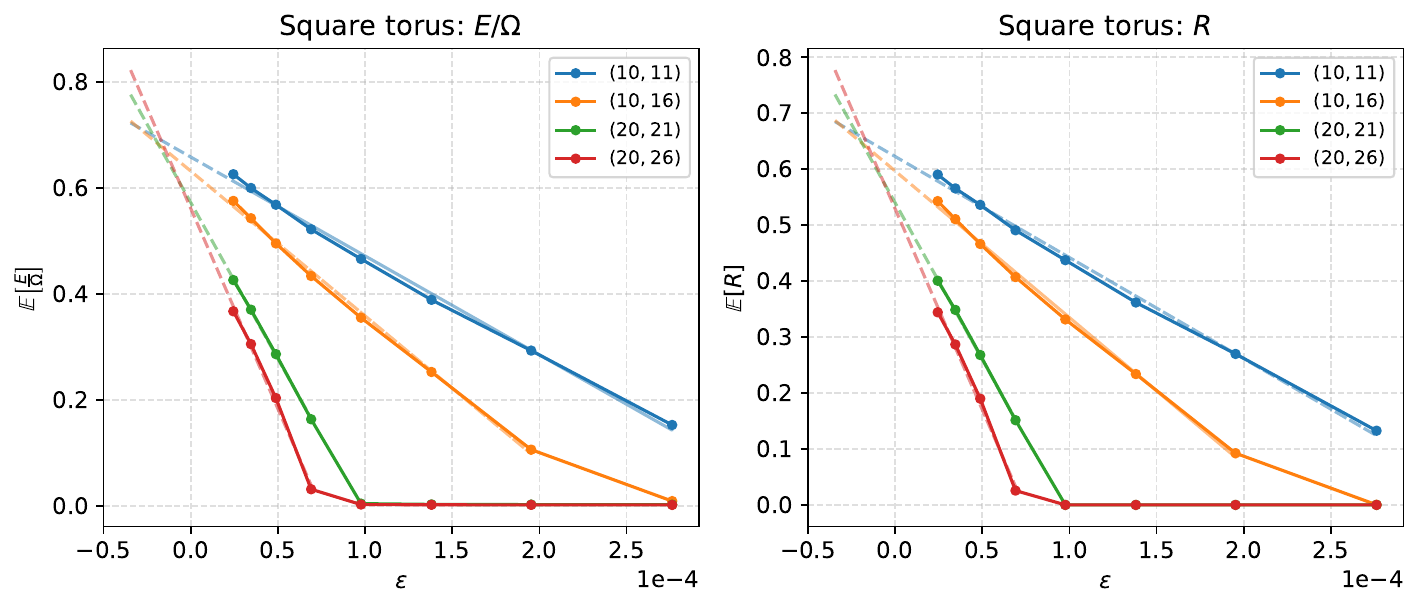}
    \caption{Energy-to-enstrophy ratio $\mathbb{E}[E/\Omega]$ (left) and fraction $R$ of enstrophy in the lowest Fourier modes (right) as functions of~$\varepsilon$ on the square torus ($N=128$, $B_0=8000$). Dashed lines show linear fits to the nonzero data, extrapolated to $\varepsilon=0$. All curves increase monotonically, and the extrapolated limits cluster within a narrow band (cf.\ Table~\ref{tab:extrapolation}).}
    \label{fig:square_EOmegaR}
\end{figure}

Figure~\ref{fig:square_EOmegaR} shows the energy-to-enstrophy ratio $\mathbb{E}[E/\Omega]$ and the fraction $\mathbb{E}[R]$ of enstrophy in the lowest Fourier modes, both plotted as functions of~$\varepsilon$ for the four forcing shells. For every shell, both quantities increase monotonically as $\varepsilon\to 0$, indicating that the stationary measure becomes increasingly concentrated on large scales. The total enstrophy remains approximately constant at $\Omega \approx B_0/2 = 4000$ throughout, consistent with the identity $\mathbb{E}[\Omega] = B_0/2$, so the growth of $E/\Omega$ is driven entirely by the upscale accumulation of energy.

At finite~$\varepsilon$ there is a clear ordering across forcing shells: shells with smaller $B_1/B_0$ condense earlier and more strongly. For $(k_{\min},k_{\max})=(10,11)$ one already observes $\mathbb{E}[E/\Omega]\approx 0.15$ at the largest~$\varepsilon$, whereas the $(20,21)$ and $(20,26)$ shells remain near the forcing-only value $\mathbb{E}[E/\Omega]\approx B_0/B_1\approx 0.002$, which corresponds to energy and enstrophy being co-located at the forcing scales with no nonlinear transfer, until $\varepsilon$ drops below $\approx 10^{-4}$, after which a sharp onset of condensation occurs. This onset should not be confused with a phase transition of the type studied in \cite{Linkmann_Hohmann_Eckhardt_2020}, where the forcing magnitude is varied at fixed viscosity.

\subsubsection{Extrapolation to the inviscid limit}
\label{sec:extrapolation}

Since the smallest simulated~$\varepsilon$ is still finite, the data cannot directly resolve the $\varepsilon\to 0$ limit. We therefore perform a linear extrapolation of each curve to $\varepsilon=0$, fitting the last nonzero data points (those with $\mathbb{E}[E/\Omega] > 0.01$) to a line. The extrapolated values are collected in Table~\ref{tab:extrapolation}.

\begin{table}[ht!]
    \centering
    \caption{Extrapolation of $\mathbb{E}[E/\Omega]$ and $\mathbb{E}[R]$ to $\varepsilon=0$ on the square torus. The effective spectral value $B_1/B_0$ measures the mean wavenumber of the forcing. The spread across shells shrinks from a ratio of $\approx 1.7$ at the smallest~$\varepsilon$ to $\approx 1.17$ after extrapolation.}
    \label{tab:extrapolation}
    \begin{tabular}{cccccc}
        \hline
        $(k_{\min},k_{\max})$ & $B_1/B_0$ & $E/\Omega\big|_{\varepsilon_8}$ & $E/\Omega\big|_{\mathrm{extr}}$ & $R\big|_{\varepsilon_8}$ & $R\big|_{\mathrm{extr}}$ \\
        \hline
        $(10,11)$ & $134$ & $0.626$ & $0.659$ & $0.590$ & $0.623$ \\
        $(10,16)$ & $220$ & $0.576$ & $0.632$ & $0.543$ & $0.597$ \\
        $(20,21)$ & $534$ & $0.426$ & $0.573$ & $0.401$ & $0.540$ \\
        $(20,26)$ & $687$ & $0.367$ & $0.562$ & $0.344$ & $0.529$ \\
        \hline
    \end{tabular}
\end{table}

At the smallest simulated viscosity ($\varepsilon_8 \approx 2.4\times10^{-5}$) the spread of $\mathbb{E}[E/\Omega]$ across shells is substantial: the ratio between the largest and smallest values is $0.626/0.367 \approx 1.7$. After extrapolation, however, the spread narrows considerably to $0.659/0.562 \approx 1.17$, and the same effect is observed for~$R$ (ratio $\approx 1.18$). This suggests that the limiting values as $\varepsilon\to 0$ may be \emph{independent} of the forcing shell, while the forcing location governs only the \emph{rate} of convergence.

This interpretation is consistent with the Sznitman--Widmayer condensation bound \cite{SznitmanWidmayer2026a, SznitmanWidmayer2026b}. In their framework, $U_0$ and $V_0$ denote twice the enstrophy and energy of the stationary state, respectively, so that $U_0 = 2\Omega$ and $V_0 = 2E$, and $\mathbb{E}[U_0-V_0]/\mathbb{E}[U_0] = 1 - \mathbb{E}[E]/\mathbb{E}[\Omega]$. Their bound (Theorem~5.1 in \cite{SznitmanWidmayer2026a}) states that for any $\ell_0 \in \{3,\dots,N\}$,
\begin{equation*}
    \frac{\mathbb{E}[U_0-V_0]}{\mathbb{E}[U_0]} \leq \frac{B_1/B_0 - 1}{\lambda_{\ell_0} - 1} + \frac{\lambda_3}{\lambda_3 - 1}\,\frac{\ell_0}{N - \ell_0},
\end{equation*}
where $\lambda_{\ell_0}$ is the $\ell_0$-th eigenvalue of the Laplacian (in increasing order, with multiplicity equal to two in their Galerkin space) and $N$ is the total number of Galerkin modes. In Section~\ref{sec:variances} we will see that the variances of $E$ and $\Omega$ tend to zero in the limit of $\varepsilon \to 0$ and therefore $\mathbb{E}[E]/\mathbb{E}[\Omega] \approx \mathbb{E}[E/\Omega]$. The first term is small when $B_1/B_0 \ll \lambda_{\ell_0}$, i.e.\ when the effective spectral value of the forcing is well below the reference eigenvalue; the second term is small when $\ell_0 \ll N$.  In our setting the forcing wavenumbers are $|\mathbf{k}|_\infty \in \{10,\dots,26\}$, giving $B_1/B_0 \in [134, 687]$, while the highest resolved eigenvalue is $\sim (\pi N/2)^2 \approx 1.3\times 10^4$.  Choosing $\ell_0$ corresponding to an eigenvalue well above the forcing band but still small relative to~$N$ makes both terms modest, and the bound predicts that forcings with smaller $B_1/B_0$ condense more strongly. This is in agreement with the earlier onset seen in Figure~\ref{fig:square_EOmegaR}.  Crucially, the bound controls only the distance from the condensed regime at finite~$\varepsilon$ and does not predict different limiting values as $\varepsilon\to 0$.  The numerical extrapolation supports the view that the limit is forcing-independent, with $B_1/B_0$ controlling the rate of approach.

We note that the extrapolation for the $(20,21)$ and $(20,26)$ shells relies on only four nonzero data points, making these estimates less certain than those for the lower shells. Moreover, the values are still slowly increasing at $\varepsilon_8$ (the relative standard deviation in the last quarter of the time series is $2$--$3\%$ for these cases), so longer simulations at smaller~$\varepsilon$ would be needed to confirm the extrapolated limits.

\subsubsection{Spectral distribution and variances} \label{sec:variances}

\begin{figure}[ht!]
    \centering
    \includegraphics[width=\textwidth]{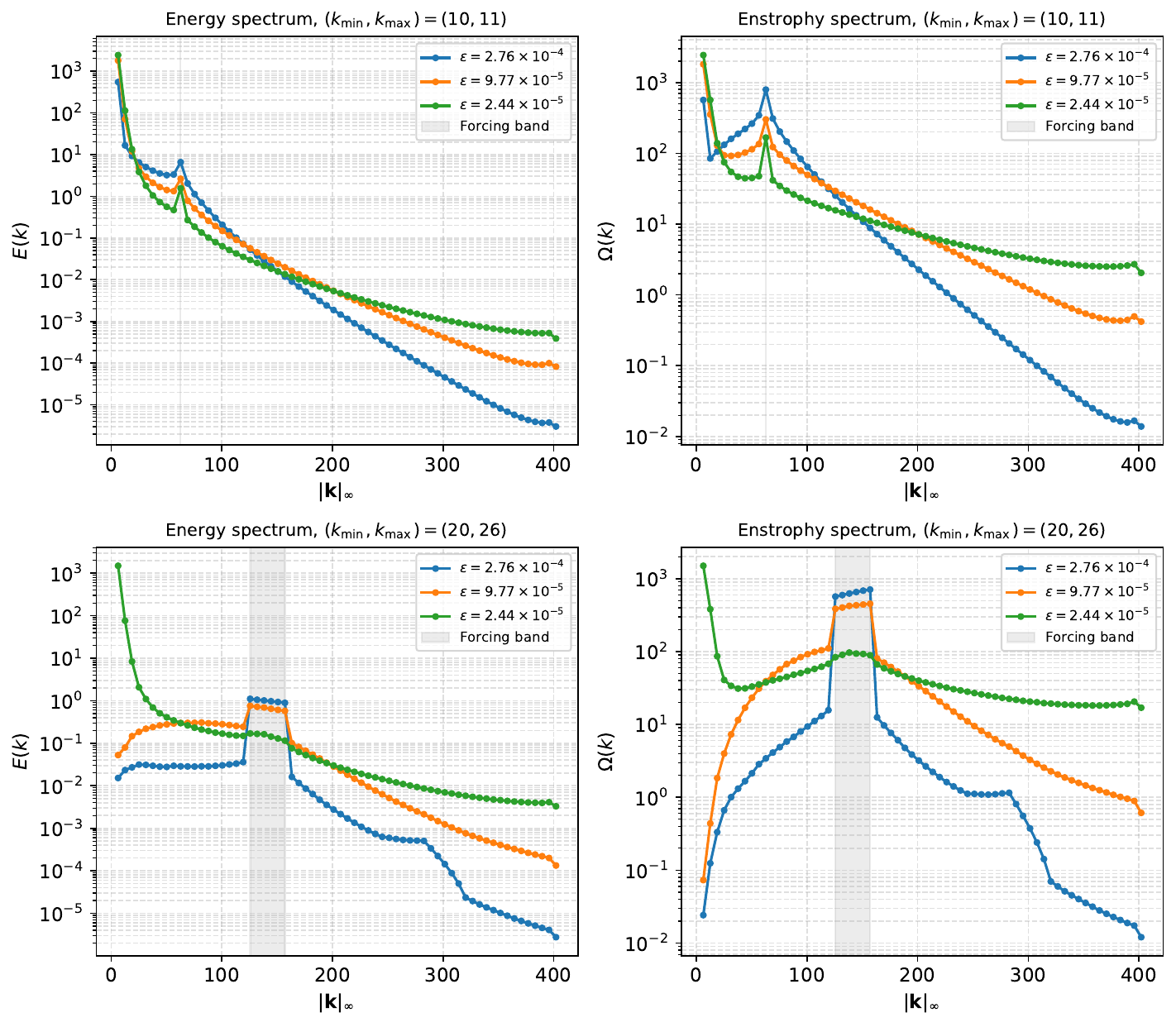}
    \caption{Radial energy (left) and enstrophy (right) spectra for $(k_{\min},k_{\max})=(10,11)$ (top) and $(20,26)$ (bottom) on the square torus at three values of~$\varepsilon$. The horizontal axis labels the shell index $|\mathbf{k}|_\infty$; the shaded band indicates the forcing shell. As $\varepsilon$ decreases, energy accumulates in the lowest modes while the enstrophy spectrum remains more broadly distributed.}
    \label{fig:spectra}
\end{figure}

Figure~\ref{fig:spectra} displays the radial energy and enstrophy spectra for the $(10,11)$ and $(20,26)$ shells at three representative values of~$\varepsilon$. At the smallest~$\varepsilon$, approximately $95\%$ of the total energy resides in the lowest wavenumber shell ($|\mathbf{k}|_\infty=1$), while only $59\%$ of the enstrophy resides in the modes with $|\mathbf{k}|_\infty=1$. These two percentages jointly illustrate that energy concentrates far more strongly than enstrophy.

\begin{figure}[ht!]
    \centering
    \includegraphics[width=\textwidth]{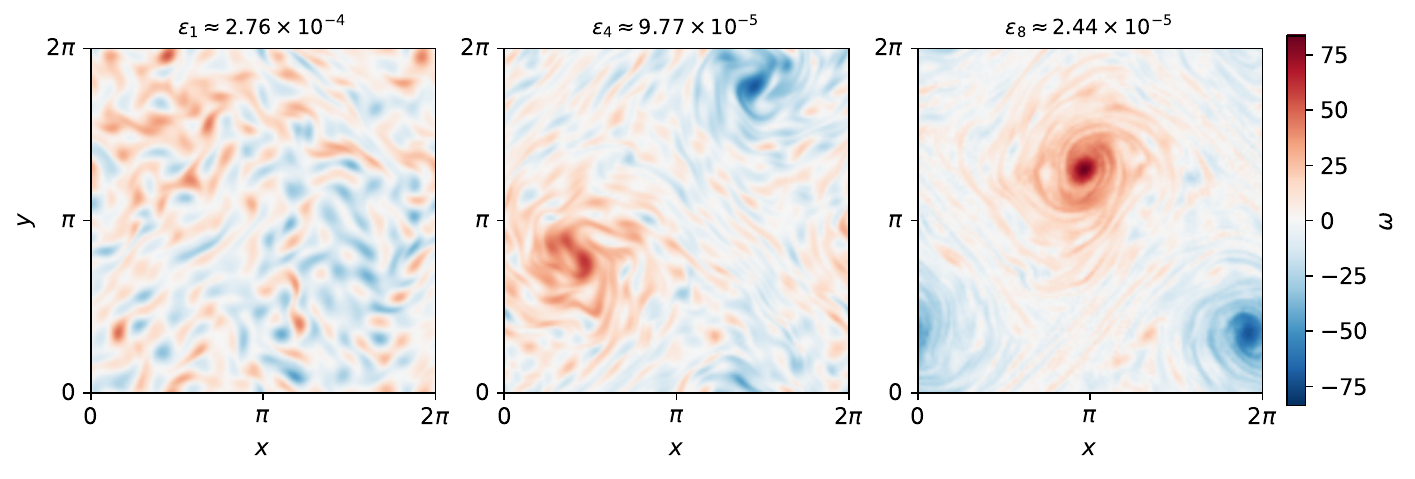}
    \caption{Vorticity field for the $(10,11)$ shell on the square torus at three values of~$\varepsilon$, from a single full-resolution ($128\times128$) snapshot at the end of each simulation. As $\varepsilon\to 0$, the field condenses into a coherent dipole structure.}
    \label{fig:flow_fields}
\end{figure}

Figure~\ref{fig:flow_fields} visualizes the vorticity field for the statistical steady-state of the $(10,11)$ shell at three values of~$\varepsilon$. As $\varepsilon$ decreases, the field transitions from a disordered state dominated by forcing-scale fluctuations to a large-scale coherent dipole, consistent with concentration of energy in the lowest Fourier modes.

\begin{figure}[ht!]
    \centering
    \includegraphics[width=\textwidth]{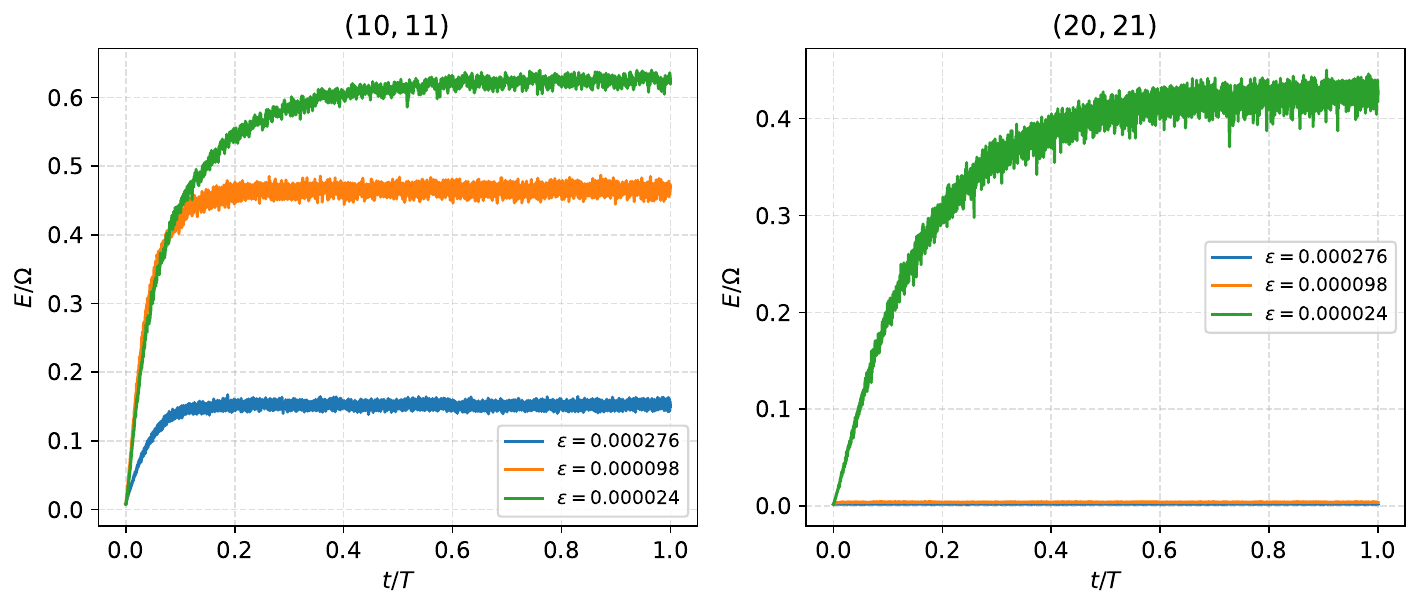}
    \caption{Time evolution of $E/\Omega$ on the square torus for forcing shells $(10,11)$ (left) and $(20,21)$ (right), each at three values of~$\varepsilon$. Time is normalized by the total simulation time $T = 4\pi^2 \cdot 3000$.}
    \label{fig:time_evolution}
\end{figure}

Figure~\ref{fig:time_evolution} shows the time evolution of $E/\Omega$ for two representative forcing shells at three values of~$\varepsilon$. For the $(10,11)$ shell, the observable reaches a plateau relatively quickly and fluctuates around a constant value for the remainder of the simulation. For the $(20,21)$ shell, convergence is slower, particularly at the smallest~$\varepsilon$, where $E/\Omega$ is still slowly increasing at the end of the simulation. Nevertheless, the relative standard deviation in the last quarter of the time series remains below $3\%$ in all cases, indicating that approximate statistical stationarity has been achieved.

\begin{figure}[ht!]
    \centering
    \includegraphics[width=\textwidth]{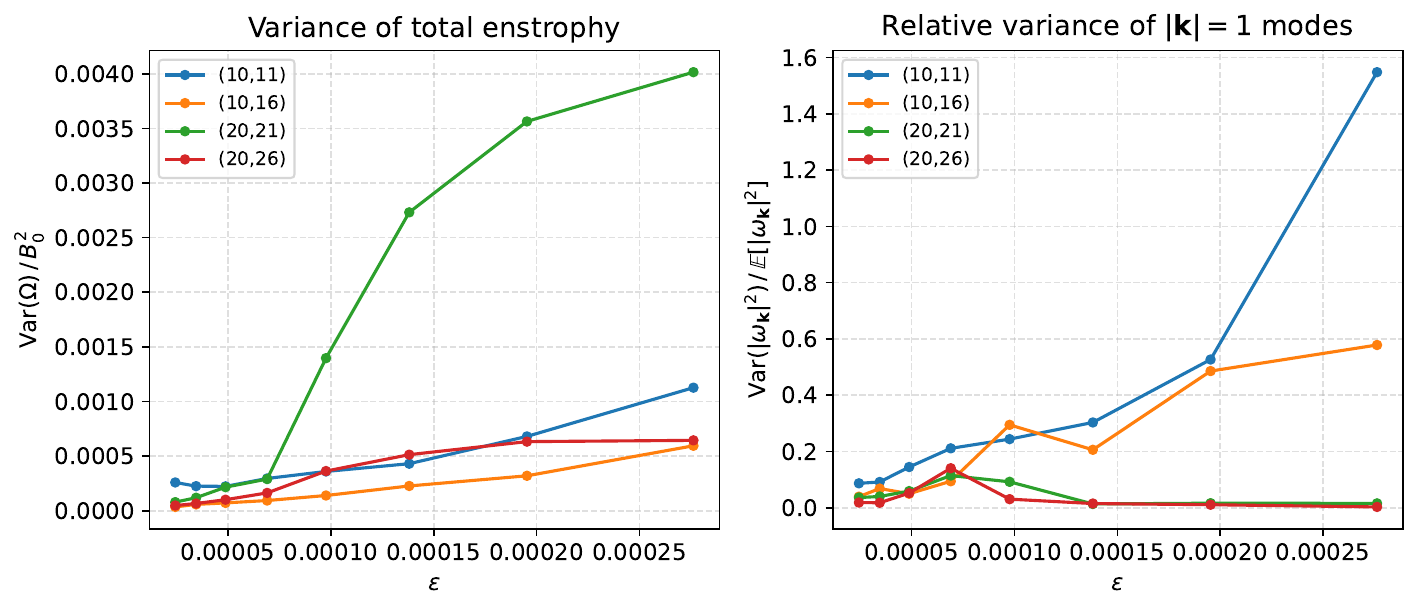}
    \caption{Normalized variances on the square torus as functions of~$\varepsilon$. Left: $\mathrm{Var}(\Omega)/B_0^2$ for all four forcing shells. Right: relative variance $\mathrm{Var}\!\bigl(\sum_{|\mathbf{k}|=1}|\omega_{\mathbf{k}}|^2\bigr)\big/\mathbb{E}\!\bigl[\sum_{|\mathbf{k}|=1}|\omega_{\mathbf{k}}|^2\bigr]$ for all four forcing shells. Both decrease as $\varepsilon\to 0$, suggesting reduced fluctuations in the fast-advection limit.}
    \label{fig:variances}
\end{figure}

The variances of global and low-mode quantities are shown in Figure~\ref{fig:variances}. The normalized variance $\mathrm{Var}(\Omega)/B_0^2$ decreases as $\varepsilon\to 0$ for all forcing shells, indicating that the total enstrophy becomes increasingly deterministic in the fast-advection limit. The relative variance of the lowest modes, $\mathrm{Var}\!\bigl(\sum_{|\mathbf{k}|=1}|\omega_{\mathbf{k}}|^2\bigr)\big/\mathbb{E}\!\bigl[\sum_{|\mathbf{k}|=1}|\omega_{\mathbf{k}}|^2\bigr]$, decreases from approximately $1.5$ at the largest $\varepsilon$ to below $0.1$ at the smallest for the $(10,11)$ shell, showing that while the mean amplitude of the low modes grows, their relative fluctuations shrink substantially. For the $(20,21)$ and $(20,26)$ shells, $\mathrm{Var}(\Omega)/B_0^2$ drops sharply once condensation sets in, reflecting the transition from a regime dominated by forcing-scale fluctuations to one governed by the large-scale condensate. The $(20,21)$ shell nevertheless stands out, exceeding the other shells by roughly half an order of magnitude at the largest~$\varepsilon$. This is a finite mode-count effect rather than a dynamical one: the normalization \eqref{eq:forcing} distributes the fixed injection rate $B_0$ over $N_{\mathrm{f}} = (2k_{\max}-1)^2 - (2k_{\min}-1)^2$ forced modes, so relative fluctuations of the band-integrated enstrophy scale as $N_{\mathrm{f}}^{-1/2}$. The $(20,21)$ shell forces only $N_{\mathrm{f}}=160$ modes, compared to $1080$ for $(20,26)$, whence the observed factor of approximately $\sqrt{1080/160} \approx 2.6$ between their fluctuation levels. The equally narrow $(10,11)$ shell shows comparable band-level fluctuations, but there only about a quarter of the total enstrophy resides in the forced band, the remainder being transferred to neighbouring wavenumbers and to the condensate, which dilutes $\mathrm{Var}(\Omega)$; for the $(20,21)$ shell the enstrophy remains confined to the forcing band, so its full fluctuations appear in~$\Omega$.

\FloatBarrier
\subsection{Thin Torus}

To test whether the above behaviour persists on anisotropic domains, we repeat the experiments on the thin torus $\mathbb{T} = (\mathbb{R}/\pi\mathbb{Z})\times(\mathbb{R}/2\pi\mathbb{Z})$ with the forcing strength set to $B_0 = 4000$. The halved forcing strength compensates for the reduced number of active modes on the sublattice, keeping the per-mode forcing amplitude comparable to the square-torus setup. All other parameters are kept identical.

\begin{figure}[ht!]
    \centering
    \includegraphics[width=\textwidth]{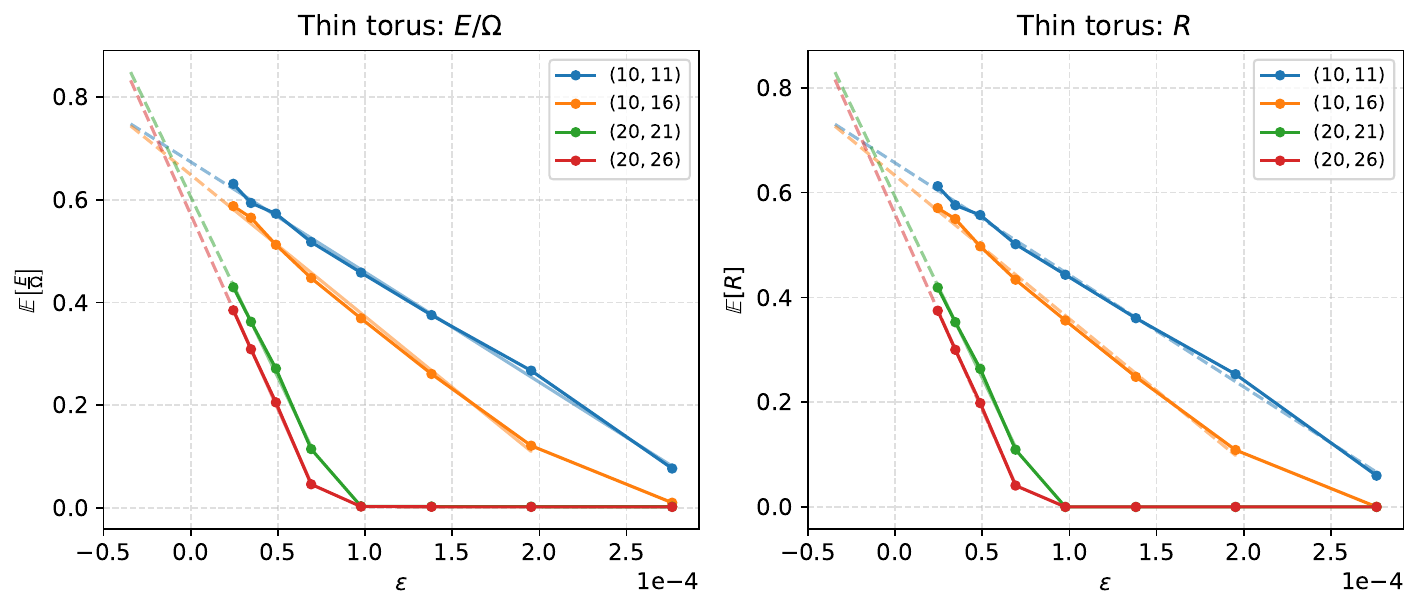}
    \caption{Same as Figure~\ref{fig:square_EOmegaR}, but for the thin torus $\mathbb{T}=(\mathbb{R}/\pi\mathbb{Z})\times(\mathbb{R}/2\pi\mathbb{Z})$ with $B_0 = 4000$. The qualitative behaviour is identical to the square torus.}
    \label{fig:thin_EOmegaR}
\end{figure}

Figure~\ref{fig:thin_EOmegaR} shows $\mathbb{E}[E/\Omega]$ and $\mathbb{E}[R]$ versus~$\varepsilon$ for the four forcing shells. The qualitative behaviour matches the square torus in every respect: both quantities increase monotonically, the ordering across shells at finite~$\varepsilon$ follows the same $B_1/B_0$ dependence, and the extrapolated values cluster together. The extrapolated $\mathbb{E}[E/\Omega]$ lies in the range $0.57$--$0.67$ (ratio $\approx 1.18$), comparable to the square-torus range $0.56$--$0.66$ (cf.\ Table~\ref{tab:extrapolation_thin}).

\begin{table}[ht!]
    \centering
    \caption{Extrapolation of $\mathbb{E}[E/\Omega]$ and $\mathbb{E}[R]$ to $\varepsilon=0$ on the thin torus. The effective spectral value $B_1/B_0$ is slightly modified relative to the square torus because the sublattice restriction removes some forced modes. The spread across shells shrinks from a ratio of $\approx 1.74$ at the smallest~$\varepsilon$ to $\approx 1.18$ after extrapolation.}
    \label{tab:extrapolation_thin}
    \begin{tabular}{cccccc}
        \hline
        $(k_{\min}, k_{\max})$ & $B_1/B_0$ & $E/\Omega\big|_{\varepsilon_8}$ & $E/\Omega\big|_{\mathrm{extr}}$ & $R\big|_{\varepsilon_8}$ & $R\big|_{\mathrm{extr}}$ \\
        \hline
        $(10,11)$ & $134$ & $0.631$ & $0.674$ & $0.612$ & $0.657$ \\
        $(10,16)$ & $212$ & $0.588$ & $0.649$ & $0.570$ & $0.632$ \\
        $(20,21)$ & $534$ & $0.430$ & $0.606$ & $0.419$ & $0.591$ \\
        $(20,26)$ & $672$ & $0.385$ & $0.571$ & $0.375$ & $0.558$ \\
        \hline
    \end{tabular}
\end{table}

\begin{figure}[ht!]
    \centering
    \includegraphics[width=\textwidth]{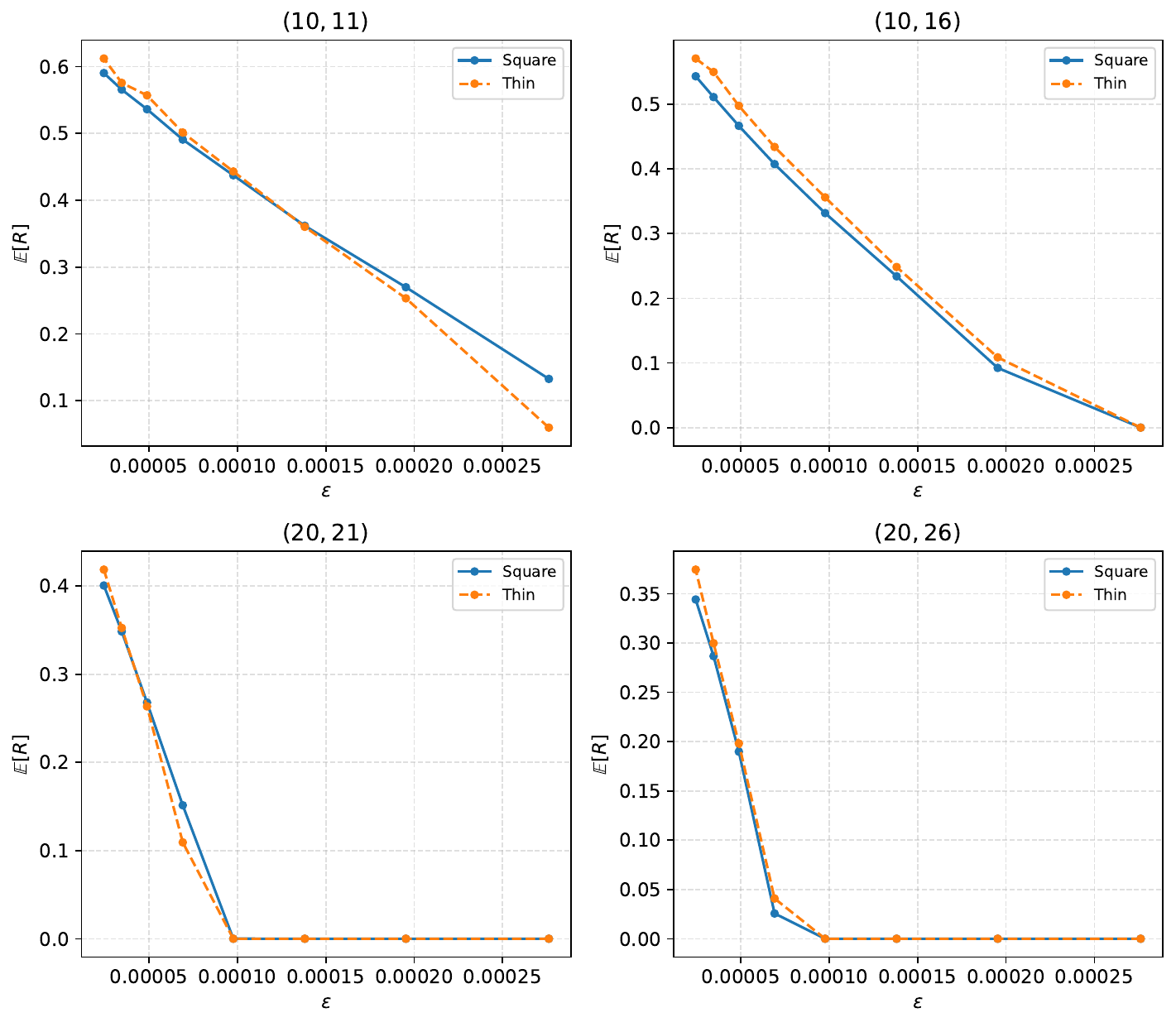}
    \caption{Comparison of $\mathbb{E}[R]$ between the square torus (solid) and thin torus (dashed) for all four forcing shells.}
    \label{fig:R_comparison}
\end{figure}

\begin{figure}[ht!]
    \centering
    \includegraphics[width=\textwidth]{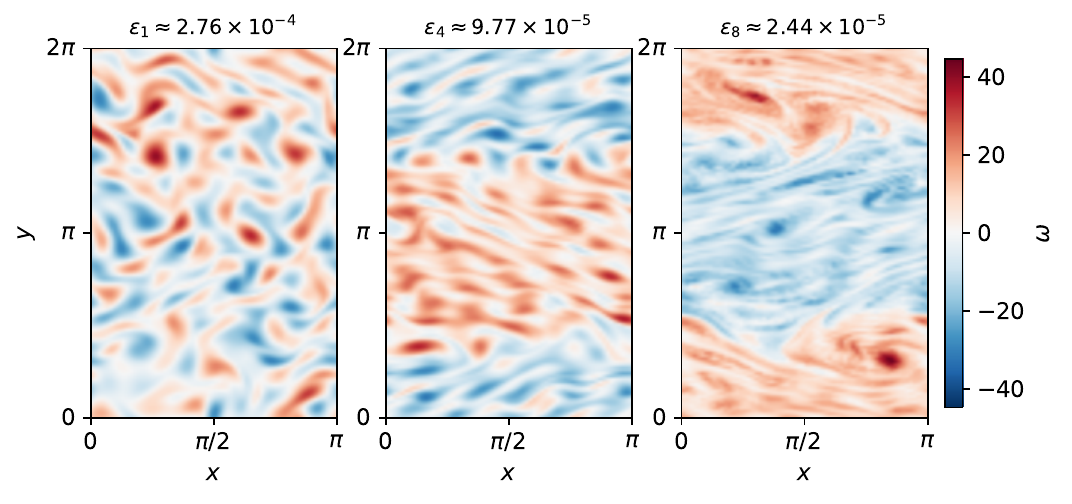}
    \caption{Vorticity field for the $(10,11)$ shell on the thin torus $\mathbb{T}=(\mathbb{R}/\pi\mathbb{Z})\times(\mathbb{R}/2\pi\mathbb{Z})$ at three values of~$\varepsilon$, from a single full-resolution ($128\times128$) snapshot at the end of each simulation. As $\varepsilon\to 0$, the field condenses into a bar-state.}
    \label{fig:thin_flow_fields}
\end{figure}

\FloatBarrier
\subsection{Discussion}

We summarise the answers to the questions posed in the introduction.

\begin{enumerate}
    \item Both $\mathbb{E}[E/\Omega]$ and $\mathbb{E}[R]$ increase monotonically as $\varepsilon\to 0$ for every forcing shell and both torus geometries. Linear extrapolation to $\varepsilon=0$ yields positive finite values, supporting convergence to a non-trivial limiting invariant measure.

    \item The stationary distribution does concentrate onto the lowest Fourier modes: at the smallest simulated~$\varepsilon$, approximately $95\%$ of the energy resides in the lowest wavenumber shell ($|\mathbf{k}|_\infty=1$). Restricting to the modes with $|\mathbf{k}|=1$ (Euclidean norm), the enstrophy fraction $R$ reaches $0.59$ on the square torus and $0.61$ on the thin torus. The energy spectrum is strongly peaked at the lowest mode, while enstrophy remains more broadly distributed, consistent with the dual-cascade picture.

    \item At finite~$\varepsilon$ the degree of concentration depends on the forcing shell, with smaller $B_1/B_0$ yielding stronger condensation. However, linear extrapolation to $\varepsilon=0$ substantially reduces the spread across shells (from a ratio of $\approx 1.7$ to $\approx 1.17$), suggesting that the limiting values may be forcing-independent.

    \item The normalized variance $\mathrm{Var}(\Omega)/B_0^2$ and the relative variance of the lowest modes both decrease as $\varepsilon\to 0$, indicating that the large-scale observables become increasingly deterministic in the fast-advection limit.

    \item The same qualitative behaviour persists on the thin torus.
\end{enumerate}

Several caveats should be noted. First, for the $(20,21)$ and $(20,26)$ shells at the smallest~$\varepsilon$, the global quantities are still slowly drifting (relative standard deviation $2$--$3\%$ in the last quarter of the time series), so these data points should be regarded as quasi-stationary rather than fully stationary; they are included in the extrapolation but contribute the largest uncertainty. The extrapolation for these shells relies on only four nonzero points and should be interpreted with caution. Second, all simulations are performed at a single resolution $N=128$. A resolution study would be needed to rule out discretization effects. However, the computational cost of simulating higher resolutions until statistical stationarity is prohibitive. Third, the linear extrapolation assumes a regular $\varepsilon\to 0$ asymptotics, which may not hold if logarithmic corrections or other singular behaviour is present. Fourth, all experiments use a single realization. The reported averages are purely temporal. We note, that under the assumption of ergodicity, temporal averages converge to ensemble averages, so the single-realization estimates are unbiased given a large enough temporal window. It is not clear, however, whether this is the case for the thin torus, as the dynamics evolve incredibly slowly along its long axis.

\section{Conclusion}

We have presented numerical evidence that stationary measures of the stochastically forced two-dimensional Navier--Stokes equations concentrate onto the lowest Fourier modes in the fast-advection limit ($\varepsilon\to 0$). The concentration is monotone in~$\varepsilon$, affects energy more strongly than enstrophy, and is accompanied by a reduction of fluctuations in both global and low-mode quantities. At finite~$\varepsilon$ the degree of concentration depends on the forcing shell through the effective spectral value $B_1/B_0$, in qualitative agreement with the condensation bound of Sznitman and Widmayer \cite{SznitmanWidmayer2026a, SznitmanWidmayer2026b}. Extrapolation to $\varepsilon=0$ suggests that the limiting values may be forcing-independent, with $B_1/B_0$ governing only the rate of convergence. The same behaviour persists on the thin torus.

The vorticity snapshots (Figure~\ref{fig:flow_fields}) reveal that the condensate takes the form of a coherent dipole, consistent with the dominance of the $|\mathbf{k}|=1$ Fourier modes. This is reminiscent of the large-scale coherent structures observed in earlier numerical studies of two-dimensional turbulence with stochastic forcing \cite{Smith_Yakhot_1993, Boffetta_Ecke_2012}, where the condensate geometry is determined by the domain shape and the structure of the lowest modes.

These findings complement the theoretical framework of \cite{SznitmanWidmayer2026a, SznitmanWidmayer2026b} by providing direct numerical evidence for the condensation phenomenon in a high-dimensional setting, and they raise the question of whether the apparent forcing-independence of the limit can be established rigorously.

\section*{Acknowledgments}

The author thanks Alain-Sol Sznitman for suggesting the experiments considered in this paper, for several helpful discussions, and for comments on the manuscript.

\null
This work was supported by a grant from the Swiss National Supercomputing Centre (CSCS) under project ID lp20.

\printbibliography

\end{document}